\documentclass[preprint,showpacs,preprintnumbers,amsmath,amssymb]{revtex4}

\usepackage{graphicx}
\usepackage{dcolumn}
\usepackage{bm}

\begin{document}

\title{Hysteretic electrical transport in BaTiO$_3$/Ba$_{1-x}$Sr$_x$TiO$_3$/Ge heterostructures}

\author{J. H. Ngai}
 \affiliation{Department of Applied Physics, and Center for Research on Interface Structures and Phenomena, Yale University, 15 Prospect Street, New Haven, CT  06520-8284, USA}
\author{D. P. Kumah}
 \affiliation{Department of Applied Physics, and Center for Research on Interface Structures and Phenomena, Yale University, 15 Prospect Street, New Haven, CT  06520-8284, USA}
\author{C. H. Ahn}%
 \affiliation{Department of Applied Physics, and Center for Research on Interface Structures and Phenomena, Yale University, 15 Prospect Street, New Haven, CT  06520-8284, USA}
\affiliation{Department of Mechanical Engineering and Materials Science, Yale University, 10 Hillhouse Avenue, New Haven, CT  06520-8267, USA}
\author{F. J. Walker}
 \affiliation{Department of Applied Physics, and Center for Research on Interface Structures and Phenomena, Yale University, 15 Prospect Street, New Haven, CT  06520-8284, USA}

\date{\today}

\begin{abstract}

We present electrical transport measurements of heterostructures comprised of BaTiO$_3$ and Ba$_{1-x}$Sr$_x$TiO$_3$ epitaxially grown on Ge. The Sr-alloying imparts compressive strain to the BaTiO$_3$, which enables the thermal expansion mismatch between BaTiO$_3$ and Ge to be overcome to achieve $c$-axis oriented growth. The conduction bands of BaTiO$_3$ and Ba$_{1-x}$Sr$_x$TiO$_3$ are nearly aligned with the conduction band of Ge, which facilitates electron transport. Electrical transport measurements through the dielectric stack exhibit rectifying behavior and hysteresis, where the latter is consistent with ferroelectric switching.   

\end{abstract}

\maketitle 

Recent advancements in epitaxial growth have enabled single crystalline ferroelectrics to be integrated with conventional semiconductors, opening a pathway to exploit the polarization of the former in semiconducting based devices. Electrically coupling the re-orientable polarization of a ferroelectric to a semiconducting channel remains a long standing challenge in materials research. Single crystalline ferroelectrics on semiconductors are ideal for such applications, since they offer superior material characteristics in comparison to polycrystalline materials \cite{Lin, Demkov}. Ferroelectric - semiconductor heterojunctions were originally conceived to operate as capacitors, where the polarization could be utilized to maintain accumulation or depletion in the semiconducting electrode\cite{Dawber}. In such an application, DC charge transport through the ferroelectric gate stack is inhibited. However, ferroelectric thin films also exhibit semiconducting properties, which could provide additional functionality in ferroelectric - semiconductor heterostructures \cite{Zubko,Bratkovsky,Morozovska,Alpay}. In conjunction with the re-orientable polarization, the semiconducting properties of thin film ferroelectrics could potentially be utilized to control charge transport through ferroelectric-semiconductor heterojunctions.

\begin{figure}[t]
\centering
\includegraphics[width=6.5cm] {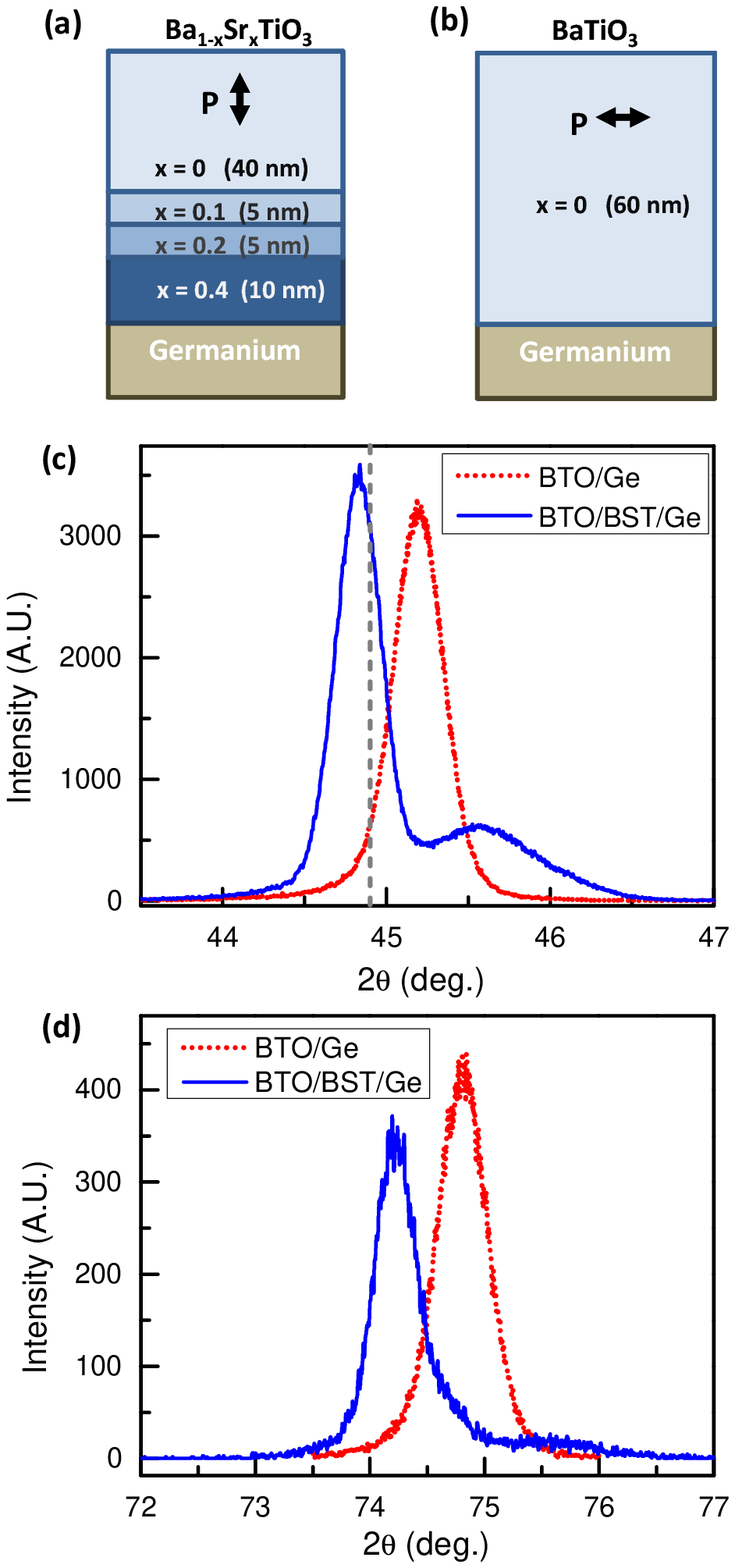}
\caption{\label{XRD_Fig1} Schematic and structural characterization of Ba$_{1-x}$Sr$_x$TiO$_3$ heterostructures grown on Ge.(a) Sr alloying near the Ba$_{1-x}$Sr$_x$TiO$_3$/Ge interface provides compressive strain to the BaTiO$_3$ top layer, enabling $c$-axis growth and a polarization that points predominantly out-of-plane. (b) Growth of BaTiO$_3$ directly on Ge results in $a$-axis growth with a polarization that lies predominantly in-plane. (c) X-ray diffraction intensity versus 2$\theta$ for the Sr-alloyed heterostructure (blue solid) shown in (a). The larger (smaller) peak at 2$\theta$ = 44.8$^0$ (46.1$^0$) arises from the BaTiO$_3$ (Sr alloyed) layer. For comparison, (002) of bulk BaTiO$_3$ 2$\theta$ = 44.9$^0$ (vertical grey dashed) is shown. In the absence of Sr-alloying, BaTiO$_3$ grown on Ge is $a$-axis oriented, as shown by the (200) diffraction peak of a 60 nm thick BaTiO$_3$ film on Ge (red dotted). (d) (103) peak of the Sr-alloyed heterostructure (blue solid) and (301) peak of the non-alloyed BaTiO$_3$ on Ge (red dotted). }
\end{figure}
 
In this letter, we introduce mobile carriers into a ferroelectric-semiconductor heterostructure and exploit the re-orientable polarization to modulate the Schottky barrier to a semiconductor. Our heterostructures are comprised of BaTiO$_3$ (BTO) and Ba$_{1-x}$Sr$_x$TiO$_3$ (BST) grown epitaxially on a Ge wafer, with Pt serving as the counter electrode. X-ray photoemission spectroscopy measurements indicate that the conduction bands of Ge and BTO are nearly aligned, thus enabling charge transport through the heterojunction interface. Transport measurements of the current through the asymmetric capacitors reveals rectifying behavior and hysteresis. The former arises from the presence (absence) of a barrier at the metal-ferroelectric (ferroelectric-semiconductor) interface for charge transport, whereas the latter can be attributed to switching of the ferroelectric polarization.

The BTO/BST heterostructures were grown using reactive molecular beam epitaxy (MBE)\cite{McKee_Science,JWReinerAdvMat}. The (100)-oriented p-type ($\rho\sim$0.018 $\Omega$cm) Ge wafers were cleaned by dipping in a 30:1 solution of H$_2$O:HF for 60 s, followed by exposure to UV light for 30 s. The wafers were then introduced into the ultra-high vacuum MBE chamber and heated to $\sim$ 500 $^0C$ to remove the native GeO$_2$ layer. A clean, dimerized Ge surface was confirmed by the appearance of a 2 $\times$ 1 reconstruction along the [11] direction of the Ge surface in the RHEED pattern. A half monolayer mix of 60:40 Ba:Sr was initially deposited on the clean Ge surface at $\sim$ 450$^0$C \cite{McKee}. The wafer was then cooled to room temperature where an additional half monolayer of 60:40 Ba:Sr was deposited, followed by co-deposition of 2 monolayers each of 60:40 BaO:SrO and TiO$_2$. The wafer was then heated to $\sim$ 500$^0$C in high-vacuum where the amorphous BaO/SrO and TiO$_2$ layers reacted to form 2 unit cells of crystalline Ba$_{0.6}$Sr$_{0.4}$TiO$_3$. Additional BST layers, described below, were subsequently grown on top of the crystallized 2 unit cells at a substrate temperature of $\sim$ 570$^0$C and in a background O$_2$ pressure of 3 $\times$ 10$^{-7}$ Torr.  Following growth, the heterostructures were annealed in oxygen plasma at 390$^0$C for 30 minutes to minimize the presence of residual oxygen vacancies. 

Our epitaxial BTO/BST/Ge heterostructures were designed to exhibit $c$-axis orientation and be sufficiently thick to offset the effects of depolarization fields \cite{LinesGlass,Tilley,Littlewood}. Depolarization fields are strong in ferroelectric/semiconductor heterojunctions due to the large screening length associated with semiconducting materials. For BTO grown on Ge, film thicknesses exceeding 40 nm are required to offset depolarization fields and maintain ferroelectric behavior \cite{JWReiner}. However, maintaining $c$-axis oriented growth for thicknesses exceeding 40 nm is challenging due to the thermal expansion mismatch between BTO and Ge. Epitaxial BTO films grown on Ge relax for thicknesses exceeding 10 nm due to the difference in lattice constants at typical growth temperatures. As relaxed films are cooled through the Curie temperature, tensile strain is imparted to the BTO from the Ge substrate, since the former has a much larger coefficient of thermal expansion than the latter. Consequently, BTO grown on Ge has $a$-axis orientation, associated with a ferroelectric polarization that lies predominantly in the plane of the film. To achieve thick $c$-axis oriented films with an out-of-plane polarization, we utilize Sr alloying to impart compressive strain to BTO, as shown in Fig.\ref{XRD_Fig1}(a)\cite{Schlom}. Our heterostructures are comprised of 4 layers, with increasing Sr content $x$ for the layers approaching the Ge substrate. The in-plane lattice constants of the BST layers near the interface are reduced with increased Sr alloying, thus compressive strain is imparted to the BTO grown on top, resulting in $c$-axis oriented films.  

\begin{figure}[t]
\centering
\includegraphics[width=7.5cm] {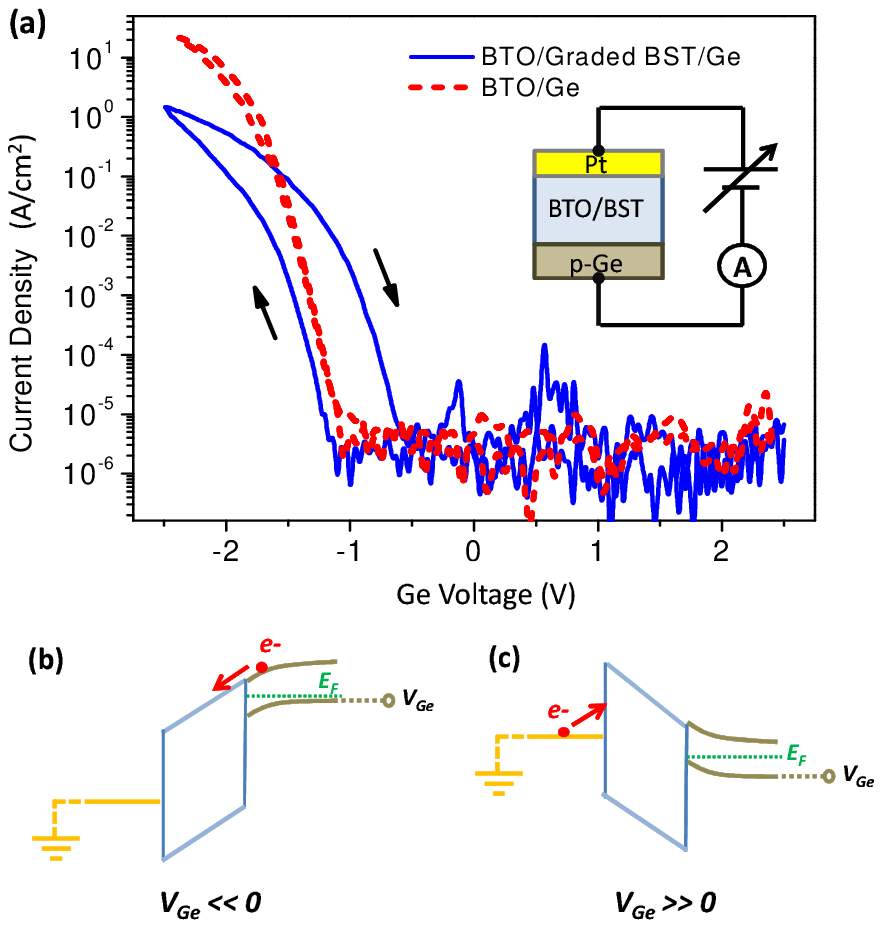}
\caption{\label{Hysteresis_Fig2} (a) Current-Voltage $I-V$ characteristics of a BTO/BST/Ge heterostructure (blue) and a BTO/Ge heterostructure (red dashed). The former (latter) has $c$-axis ($a$-axis) orientation and exhibits (does not exhibit) hysteresis. (b) For $V_{Ge}$ $<$ -1 V, charge transport is enabled due to alignment of BST and Ge conduction bands. (c) For $V_{Ge}$ $>$ -1 V, charge transport is inhibited by the barrier between Pt and the BaTiO$_3$ conduction band.}
\end{figure}
X-ray diffraction measurements taken on our heterostructures confirm $c$-axis oriented growth. Figure \ref{XRD_Fig1}(c) shows x-ray intensity versus 2$\theta$ for a typcial BTO/BST/Ge heterostructure (blue curve). The smaller (002) peak at 2$\theta$ = 45.6$^0$ arises from the 20 nm thick Sr alloyed layers near the interface, while the larger (002) peak located at 2$\theta$ = 44.75$^0$ is associated with the 40 nm thick BTO top layer. For comparison, the (002) associated with bulk BTO is indicated by the dashed grey line. We note that the compressive strain induced by the Sr alloying has enhanced the out-of-plane lattice constant of our BTO films to exceed the bulk value. Figure \ref{XRD_Fig1}(c) also shows x-ray intensity data for a 60 nm thick BTO/Ge film without the Sr alloyed layers (red dotted), schematically illustrated in Fig.\ref{XRD_Fig1}(b). The absence of Sr-alloying results in $a$-axis oriented films with predominantly in-plane polarization. Finally, Fig.\ref{XRD_Fig1}(d) shows the (103) (blue solid) and (301) (red dotted) peaks for the BTO/BST/Ge and BTO/Ge heterostructures, respectively. Analysis of the x-ray data for the BTO/BST/Ge heterostructure indicates the out-of-plane (in-plane) lattice constant is 4.040 \AA (3.987 \AA).

Previous studies have shown that the band gaps of BTO and SrTiO$_3$ are virtually identical, and that their conduction bands are nearly aligned with the conduction bands of Si and Ge \cite{Cardona,FAmy,Robertson}. The near alignment of conduction bands at interfaces between BTO/BST and Ge facilitates charge transport and suggests that rectifying behavior should be exhibited by such heterojunctions, provided a counter electrode forming an interface with finite barrier for charge transport can be created on the BTO surface.

To explore the possibility of diode-like transport, Pt electrodes 28 $\mu$m in diameter were deposited by electron beam evaporation through a shadow mask on the BTO surface, and 2-point electrical transport measurements were performed in the geometry illustrated in the inset of Fig.\ref{Hysteresis_Fig2}(a). Representative current-voltage characteristics of a Pt/BTO/BST/Ge heterostructure are shown in Fig.\ref{Hysteresis_Fig2}(a). For $c$-axis oriented heterostructures, rectifying behavior and hysteresis is observed (blue solid). Transport characteristics of an $a$-axis oriented Pt/BTO/Ge heterostructure are shown for comparison (red dashed), which also exhibits rectifying behavior but negligible hysteresis. The rectifying behavior arises from the asymmetry of the metal/ferroelectric/semiconductor heterojunction, namely the presence (absence) of a potential barrier at the Pt/BTO (BST/Ge) interface. For voltages applied to the Ge wafer that are $V_{Ge}$ $<$ -1 V, electrons flow from Ge to Pt electrodes via the BST conduction band\cite{Minority_Generation}, as illustrated in Fig.\ref{Hysteresis_Fig2}(b). In contrast, for $V_{Ge}$ $>$ -1 V, a barrier arising from the workfunction of Pt inhibits holes from entering the BTO valence band, as illustrated in Fig.\ref{Hysteresis_Fig2}(c).

\begin{figure}[t]
\centering
\includegraphics[width=7.5cm] {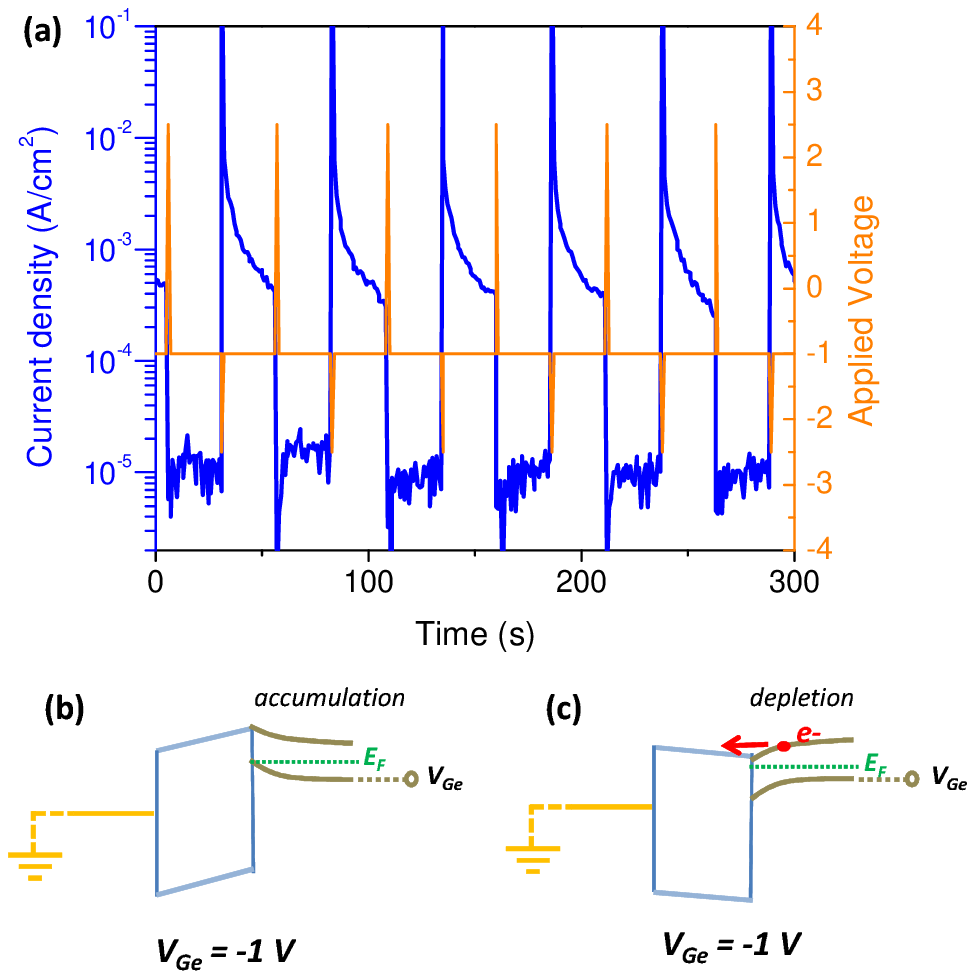}
\caption{\label{pulsed_measurements_Fig3} (a) Transport current through Pt/BTO/BST/Ge heterojunction (blue) versus time as a function of applied voltage (orange). (b) and (c) Schematic illustrating two orientations for the polarization and the associated barrier heights for carrier transport for $V_{Ge}$ = -1 V. The state of near inversion in (c) promotes transport of minority carriers through the ferroelectric stack. In contrast, the state of accumulation in (b) forms a barrier for the transport of minority carriers through the ferroelectric stack. }
\end{figure}

The hysteresis in the transport current for $c$-axis oriented heterostructures exhibits characteristics that are consistent with ferroelectric switching. First, the hysteresis is present (absent) for $c$-axis ($a$-axis) oriented heterostructures, consistent with switching (non-switching) under electric fields that are applied co-linear with (perpendicular to) the direction of polarization. Second, the magnitude of the transport current can be switched between high and low regimes with $\sim$ ms duration voltage pulses centered about $V_{Ge}$ = -1 V, as shown in Fig.\ref{pulsed_measurements_Fig3}(a). Thus the hysteresis cannot be explained by effects associated with the drift of mobile ions, since ions within dielectric stacks typically do not respond to voltage sweeps that are $\geq$ 5 V/s. \cite{TPMa} Third, the ``clockwise'' direction of the hysteresis shown in Fig.\ref{Hysteresis_Fig2}(a) is consistent with switching of the polarization, as illustrated schematically in Fig.\ref{pulsed_measurements_Fig3}(b) and (c). As $V_{Ge}$ is swept from -2.5 V to -1 V, the polarization of the BTO causes the Ge to be near inversion, as shown in Fig.\ref{pulsed_measurements_Fig3}(c). The state of near inversion, implies the absence of a Schottky barrier for minority carriers to transfer from the Ge to the BTO/BST heterostructure. In contrast, as $V_{Ge}$ is swept from +2.5 V to -1 V, the direction of the polarization is reversed, thus the Ge is in accumulation. In this scenario, a Schottky barrier prevents minority carriers from transferring from the Ge to the BTO/BST.       

Carriers injected into the Ge through the BTO/BST barrier enables the interplay between ferroelectric and semiconducting properties to be explored. It is interesting to note that the change in transport current in response to pulsed voltages initially spans over 3 orders of magnitude but decays with time, as shown in Fig.\ref{pulsed_measurements_Fig3}(a). The decay of the transport current with time could be associated with weakening of the polarization due to the presence of space charge\cite{Zubko}. Alternatively, the decay could arise from depolarization fields weakening the ferroelectricity\cite{JWReiner}. It should be noted that a constant field within the BTO/BST dielectric stack is shown in Fig.\ref{pulsed_measurements_Fig3}(b) and (c). A more rigorous description should consider gradients in polarization and fields due to the Sr alloying\cite{Bratkovsky,Morozovska,Alpay}. Finally, we note that the transport current effect observed in our heterostructures could potentially form the basis for a type of memory.  

In summary, we have grown and performed structural and transport characterization of epitaxial, $c$-axis oriented BTO/BST/Ge heterostructures. The thermal expansion mismatch that prevents $c$-axis oriented growth is overcome through Sr-alloying, which imparts compressive strain to the BTO. The conduction bands of Ge and BaTiO$_3$ are closely aligned. Transport measurements through the $c$-axis oriented heterostructures exhibits rectifying behavior and hysteresis in the transport current. The rectifying behavior of the Pt/BTO/BST/Ge heterostructures arises from the asymmetry in band alignments at the Pt/BTO and BST/Ge interfaces, while the hysteresis exhibits characteristics that can be attributed to switching of the ferroelectric polarization.

\begin{acknowledgements}
This work was supported by the NSF under contracts DMR-1119826 and MRSEC DMR-1309868 (CRISP). 
\end{acknowledgements}

\end{document}